# Distributed Software Evolution: a Survey

Mohammad Reza Besharati

PhD Candidate, Computer Engineering Department, Sharif University of Technology, Tehran, Iran, Corresponding Author,
besharati@ce.sharif.edu

**Abstract**

Distribution can be a feature of the software evolution process. In other words, temporally and spatially distributed teams and organizations can develop and work on a software application. The simplest case is to outsource production and employ workforce at distributed sites so that multiple distributed teams can work on a project within a parallel framework. If this distribution is global, it will be called the global software evolution/development. A higher level of distribution is defined as decentralization and decentralized software evolution, which means that software development can be independent of the initial provider. It also means that software execution is independent of the initial provider and the initial system so that the software application can easily be reused in different and new projects. However, the high-level architecture is managed within a practically centralized framework in the decentralized software evolution. Most of the large-scale open-source projects are exemplars of this level. In terms of distribution, there is a higher level of decentralized software evolution called "distributed cognition and leadership". At this level of distribution, all system levels evolve within a distributed framework, and there are no centralized points in the project network and its evolution process. Some open-source software applications are the exemplars of this last level. Not only is the distributed software evolution faced with certain challenges and opportunities to reach its goals, but it has also caused some challenges and opportunities in other fields. This paper conducts a general review of the distributed software evolution. For this purpose, the paper first addresses the importance of the distributed software evolution, and then introduces its noteworthy paradigms. The tools, requisites, outcomes, and challenges of the distributed software evolution are finally discussed.

## Formation of This Field

Ref. [5] is an old paper that can properly show the research background. Published in 1997, this paper indicates how the need for further development and evolutions of different tools such as CVS was expressed.

This paper emphasizes that many software applications are now developed by distributed teams within very distributed frameworks. In this case, a concern is to complete the software development environment at the same time as the software evolution process. In particular, file management in the development environment monitors the changes made to files over time and emphasizes the knowledge about this process and the history of these changes.

According to this paper, it should be possible to create and manage files within a distributed framework in the distributed software development environment to become aware of the history of their changes. However, the system introduced by the paper should be capable of distributed file management in the development environment in the future, something which was not feasible at the time.

## Importance of Distributed Software Evolution

In Ref. [1], Tom Mens introduced the distributed software development clearly as a challenge to software evolution among emerging paradigms and associated technologies. However, this reference pertains to the slides that he prepared in 2009. Tom Mens (2005) authored a paper of the same theme and addressed the same subject in a more general way. The importance of this reference lies in the fact that it indicates that the subject of distributed software evolution was recognized by the experts of this field such as Tom Mens. More importantly, they have mentioned certain challenges that should be handled in the future. Apparently, Tom Mens talked more about the (geographically and temporally) distributed through their coordination and synergic performance for software evolution without being aware of each other's activities. It should be mentioned that this viewpoint of software evolution is among the most basic considerations and is a principle of software evolution.

Ref. [2] was authored by Mens *et al.* in 2009. Like Ref. [1], it introduced the distributed software development as one of the emerging and challenging paradigms as opposed to software evolution.

According to Ref. [15], the Web-based environment has the potential to be the infrastructure for a global software engineering environment which can support software evolution in all steps of its lifecycle, regardless of the number of individuals involved in the software project and their stations. The easy recruitment of human resources for software projects is considered an encouragement and outcome of this process. By considering the number of people who access the Internet, it is ultimately possible to find sufficient individuals who are interested in the project goal and participate in the distributed process of software development and evolution. This paper was later validated practically by the success of SourceForge, Google-Code, and the other open-source projects on the Web.

In another point of view, software-based virtual enterprises can be discussed (*e.g.*, Apache Project). In particular, there are many individuals who are naturally willing to participate in a software project. In fact, a project will provide further participation opportunities that can attract more human resources if it grows further, expands, and become more complicated. Due to an open, distributed, and participatory approach in a virtual enterprise, human resources are recruited faster if they are needed further. In other words, virtual enterprises have a kind of self-adaptation mechanism, which is an outcome of distributed software evolution.

The facilitation of human resources recruitment and the self-adaptability of individuals and organizations to a project can improve the process agility, help show faster reactions to changes, and upgrade some managerial aspects of a project as well as its hazards. It can also affect and enhance the other aspects of a project in which individuals and an organization play parts.

Moreover, when the constraints of installation site and participation time are lifted and diminished by using distribution in the process, experts can work on multiple projects simultaneously. Therefore, the experts of an organization are employed more efficiently. It is also possible to reduce the depression of monotonous work, which is a prominent problem with software engineering. Furthermore, more projects can employ the experts. In fact, not only is the human resources recruitment facilitated in a project, but it is also possible to improve the quality of human resources.

The Web-based distribution of a software development environment can help constantly improve the software quality and the development environment through massive amounts of feedback. This is a well-known topic in open-source software.

The reuse of software products can also be improved in the distributed software evolution, especially if the Web-based infrastructure is employed. Therefore, it is easy to have a global market of software component providers, something which is practically happening today.

This paper proposed adding the necessary technology for the interaction and coordination of individuals with each other in a project as a standard component to the Web-based infrastructure. However, with the passage of time, this does not appear necessary now. In fact, the application layer of the Web architecture has the necessary capacity and sufficient flexibility to provide this technology effectively.

Ref. [17] proposed the open architecture, which is based on the decentralized software evolution—*i.e.*, an exemplar of the distributed software evolution— for the engineering of complicated systems.

It is essential for business organizations, software projects, and software developers to learn from the environment. Learnability is a determinant of success for an organization or an individual. The distributed software evolution increases the degree of parallelism in the development process, something which improves learning (*e.g.*, failures and successes are considered a source of learning. Each of the parallel procedures for a development process will end up in either failure or success; thus, increasing the degree of parallelism in a development process would be equal to increasing the learning rate of that project).

The distributed software evolution allows for the integration of the end user development into the conventional software development and evolution (based on the separation of the developer and the end user). In fact, the distributed software evolution can help involve the end users of a system (that are typically distributed for many distributed software systems) along with the other project stakeholders in the software evolution process. This means a new paradigm and an intermediate border between the end user development and the conventional software development and evolution, something which was mentioned in Ref. [18].

The distributed software development is similar to the distributed software evolution. At least, both emphasize the spatial distribution. In fact, both are based on the interaction and participation of distant individuals. Both are faced with similar challenges, *i.e.*, the involved parties are less likely to have physical interaction and exchange. Hence, it is necessary to use certain tools for facilitating the exchange, interaction, and relationship of individuals. This paper takes a theoretical look at the relationship between the involved individuals and the adoption of the proper medium for this relationship by listing the software tools that facilitate the relationship, exchange, and interaction of the involved individuals.

Quoting another paper, Ref. [20] introduced six historical motives for the globally distributed software development as below:

1) The mergers and contractors that practically relate the software development teams on a global scale
2) Software companies think about global markets and global positions to improve their businesses.
3) Increasing knowledge and approaching the market, which became global; there should be distributed sites to have up-to-date information on the distributed clients
4) Accessing the most skilled developers: They are not centralized in one location. The most skilled developers are scattered all over the world. There should be a distributed framework to employ these experts.
5) Reducing production cost: Developers are paid differently in different countries. The globally distributed development helps make decisions on developers with different wages to select the best case.
6) Reducing the "time to market": The distributed software development process does not stop even when the day turns into the night. At any time of the day, there are some individuals who turn the wheels of the project.

This paper also included the interesting terms "global software development" and "global software engineer" referring to a specific type of software development with the global participation of individuals. Ref. [21] (2001) introduced the globally distributed software development and open-source software as an emerging trend.

In the ICSE (2012), the fourth session of the Software Engineering Education Track included teaching the distributed software engineering in areas such as global software engineering and models of interaction in the distributed software development projects.

**Specific Paradigms in Distributed Software Evolution**

**Decentralized Software Evolution**
Published by Mr. Arizi and Dr. Taylor in 2003, Ref. [3] discussed the distributed software evolution and the ability to develop a software application in a way that would be independent of its initial developer. It also means to develop the software components that are independent of the initial developer and the system in which they were used for the first time; therefore, they can easily be reused in different and new projects. This has been achieved in other industries to some extent. In fact, adaptability and customizability are the solutions offered for this purpose.

The authors of this paper believe that although such techniques, principles, and methods as object-orientation, aspect-orientation, separation of concerns, design models, and modularity have improved the reuse of software products and the decentralized software evolution, their flexibility is overshadowed by the conventional methods of connecting components for implementation (*e.g.*, conventional linkages and compilations). According to the authors:

> "The compilation process solidifies the plasticity of a design."

Many papers on the distributed software evolution addressed the concern of compilation for the components that are distributed (in terms of time, location, or other aspects). This can be considered another topic in the distributed software evolution.

Moreover, the decentralized software evolution is a kind of the distributed software evolution, for it has the features of the distributed evolution (*i.e.*, distribution in time, location, *etc.*). In addition, it emphasizes the independence of evolution procedures and the lack of a center in the evolution process as much as possible.

Furthermore, compilation is only one of the concerns in the distributed software evolution. There are other concerns such as population and consistency of evolution sub-processes. Some of these concerns are introduced later.

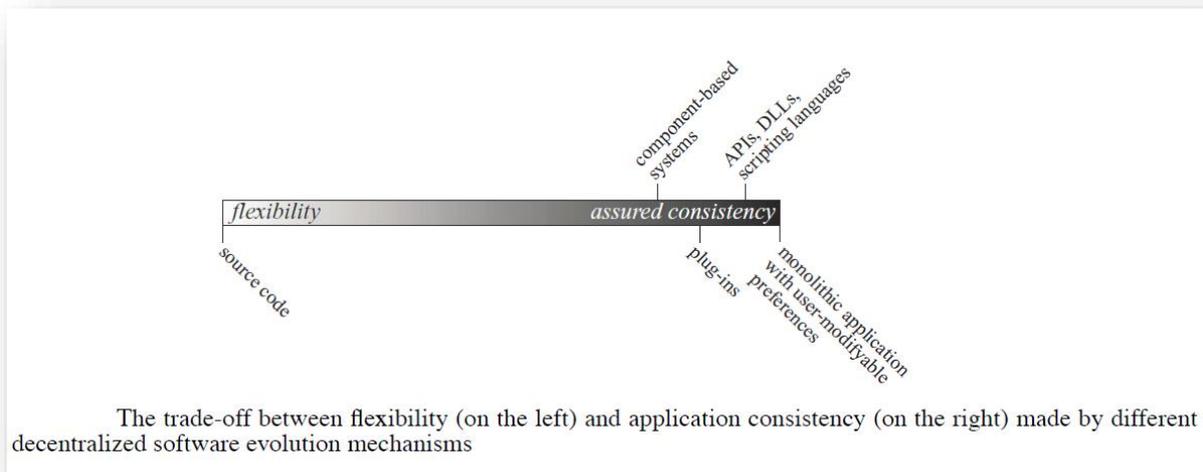

**Figure 1.** The trade-off between flexibility and application consistency in different decentralized software evolution mechanisms

Authored by Arizi and Taylor, Ref. [4] continues the way passed by the previous paper. If a system is to evolve in a way that is independent of a centralized provider, *i.e.*, having different providers that complete a part of the system within a decentralized framework, how can we guarantee the universality and integrality of the system? How can we ensure that the system components are consistent? In fact, these questions arise from the decentralized software evolution.

A doctoral dissertation submitted by Arizi, Ref. [9] addressed the decentralized software evolution, which was introduced earlier. Authored in 2000, this dissertation proposed a novel software customization approach that was improved in terms of adaptability and consistency compared with the other approaches. Called the *Open Architecture Software*, the proposed approach was based on the idea of separating the implementation layer from the high-level architecture layer and defining an "evolution administrator" responsible for coordinating these two layers. The distributed software component providers construct and create the implementation layer.

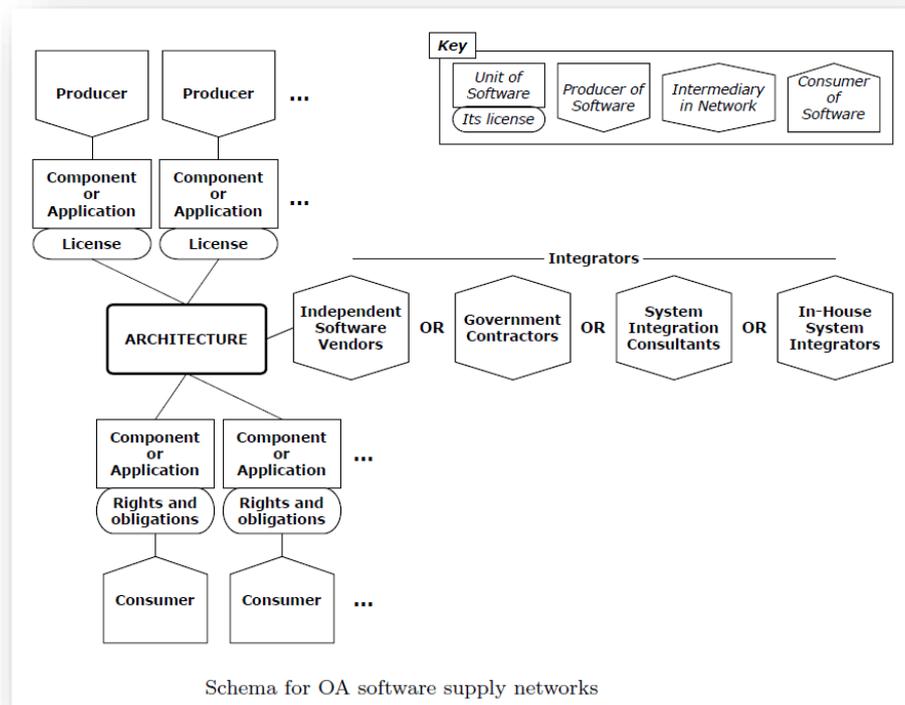

**Figure 2.** The software supply network through *Open Architecture* (Ref.: Paper 32 taking this scheme from another reference)

Ref. [13] is a master thesis (2011) that proposed a developer-oriented interface for the systems based on *Open Architecture*. This thesis aimed to facilitate the relationship between developers and the evolving system. In this regard, it took a different viewpoint from the previous references at the distributed software evolution. In fact, this thesis tried to manage system complexity to facilitate the tasks handled by developers.

The next interesting topic is ectropy software. In fact, ectropy is the opposite of entropy. In other words, it means the open-source software that evolves in a distributed process; however, its structural consistency increases gradually and does not decrease. Ref. [10] discussed a prominent defect of the distributed software evolution known as the threatened consistency of components and the need for maintaining the logical structure of architecture over time. This reference proposed a solution based the idea of separating the concern for the definition of structure and architecture and the concern for the implementation of components. Therefore, this reference and the previous one can be considered similar in terms of the approach that they adopted.

**Open-Source Software**

A category of papers on the distributed software evolution includes the papers on the open-source software which usually evolves in a distributed framework (in terms of time, location, *etc.*). Ref. [12] is a detailed survey (2010) conducted on the free open-source software. This reference discussed the "evolution of the developer society" of an open-source software application.

Ref. [11] proposed an architecture for the distributed software evolution for the applications designed for the environments set to provide distributed and even customized services for the distributed clients.

It can be concluded that this paper and other similar papers addressed the concern for distribution as a characteristic for software evolution. In other words, the systems that use a kind of distribution in their processes have a specific feature (*i.e.*, distribution), based on which their components and processes should be configured.

Ref. [19] worked on the anti-patterns and defects of open-source software development such as the defect or lack of documents, inconsistency of components, lack of coordination between developers, lack of communication between developers, lack of qualitative review of software products and its low quality, insufficient information about the team status, insufficient information about the project status among the involved individuals, lack of reporting and getting feedback from users, and failure to use their experience for the constant software improvement. This paper proposed a method for eliminating anti-patterns and defects. For this purpose, the necessary knowledge about the status and background of a software project should first be provided in a knowledge repository to enable the project manager to find a solution to the elimination of anti-patterns and defects. In the proposed method, a CVS ontology and a bug ontology are developed to determine the data structures that should be stored in the knowledge repository within the development process.

A specific category of software projects evolving within a distributed framework are Free/Libre/Open-Source applications abbreviated to FLOSS.

Based on a theoretical framework of ecological metaphor, Ref. [25] analyzed the possible presence of "edge effects" on the software development and evolution methods. In fact, the "edge effect" is an ecological terminology indicating that the biological complexities of two biomes put together will result in new biological complexities on the "edge" and encounter of these two biomes. Similarly, complex practices can emerge as the integration of "open-source software" and "proprietary software" evolution and development. Not only are these practices open-source, but they are also proprietary.

Ref. [31] addressed the necessity of analyzing the evolution of a participant population in an open-source project in addition to studying the software evolution by proposed a method for the integrated analysis of these two evolutionary procedures (open-source software evolution and the evolution of developers and participants). In fact, the software architecture affects the attraction or detraction of participants and the evolution of the participant population. At the same time, the participants of an open-source project progress the project; therefore, making changes to their groups (*e.g.*, how they join the project, how they leave the project, how they affect software transformations, how an individual's position is upgraded among other participants, *etc.*) will directly affect the process of an open-source project.

An important aspect of projects based on *Open Architecture* is the software license, which can change over time. There are different types of licenses that might change many times during the software lifecycle. This can be a research topic (Ref. [32]).

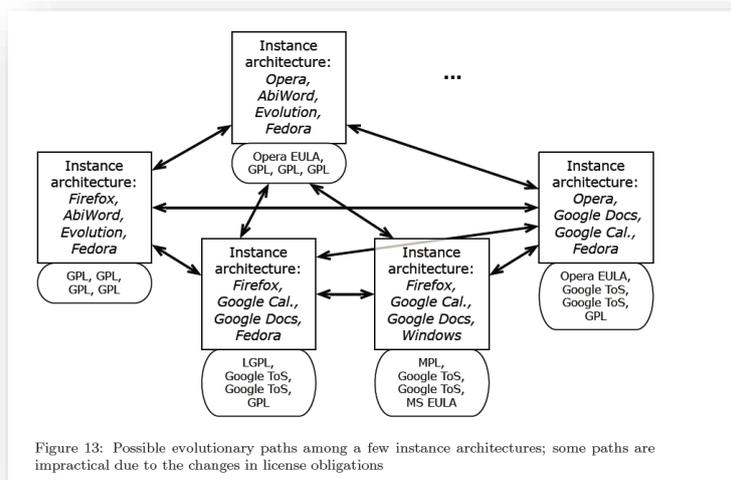

**Figure 3.** The evolution of licenses (Ref. [32])

In FLOSS projects, a problem is that some tasks are popular with participants, whereas some other tasks are not. This problem is known as the "popularity phenomenon" in this field (Ref. [33]). This kind of evolution and development process in FLOSS projects may lead to an unbalanced growth. For instance, the bugs of a specific component might be eliminated very quickly; therefore, it will be practically reliable and high-quality. However, the bugs of another component in the same system might be eliminated after a long time. Hence, the bugs are either not detected on time or left undeleted for a long time after they are detected. As a result, that component is practically unreliable and will have low quality.

Another problem with open-source projects is that the system architecture sometimes emerges after the system is developed instead of being designed in advance. This is a characteristic of complex systems. In other words, the next system status cannot be predicted nearly accurately. In fact, the next system status emerges in practice, something which indicates the process of evolution in open-source projects and the participants are practically the components of a complexity and a complex system. The next complexity emerges from their interaction. However, if the *Open Architecture* is used, such an incident should not occur seriously. In this case, the high-level architecture layer should direct the entire evolutionary process of the system.

**Miscellaneous Items**

The "distributed cognition" is another interesting topic on the common denominator of distribution and social media: That is the time when distributed individuals and entities act as a unified cognitive entity to solve a common problem (Ref. [36]).

Ref. [37] addressed this problem along with the "distributed leadership" in the environments of online creative and interactive environments, an exemplar of which is the distributed software evolution.

Ref. [38] made a subtle point. The growth rate of other IT markets (*e.g.*, portable devices, mobiles, tablets, smartphones, secondary hardware, *etc.*) depends on the growth rate of their available applications. If the applications are not developed quickly in large amounts, the use of such hardware will partly be limited. In fact, these devices will remain out of use, and the demand for these markets will practically reduce. Sometimes, it is better to develop such bottleneck products (in this case, software applications) without considering the calculated profit or loss so that the other markets will not experience recession and demand reduction. This is the exact incidence that has happened to the applications developed for mobiles, tablets, and smartphones. However, there is an intermediary solution. It is possible to achieve a high software development rate at a low cost through crowdsourcing on the platform of FLOSS and open-source paradigms. Hence, the applications developed through crowdsourcing have different roles and responsibilities from the other applications. They also have different areas of use. Although each case of crowdsourcing cannot be attributed to the abovementioned cause, many of the cases are implemented for this purpose. The concept of crowdsourcing is similar to the previously discussed concepts.

Ref. [40] analyzed the effects of the distributed software evolution. For instance, the distributed software evolution can lead to the migration of capitals and job opportunities from the leading countries in IT to the developing countries. Moreover, these socioeconomic effects bring responsibilities for the researchers of this field; thus, it is necessary to revise and expand our knowledge of the distributed software evolution by considering scientific ethics. A few questions have also arisen in this regard. Empirical studies and analyses of scientific ethics have also been requested in addition to the other research methods for answering these questions. It is possible to crowdsource all components of a software application for distributed evolution? What conditions and constraints are essential for the distributed software evolution? What control mechanisms are required? How can we determine the authenticity and health of the procedure for generating a snippet? Is it possible? Is it favorable? This paper indicates that the distributed software evolution had a significant effect on socioeconomic areas. Finally, it should be mentioned that the interesting term "distributed software engineering" was used in Ref. [16].

**Tools, Requisites, Outcomes, and Challenges**

Ref. [6] (2010) proposed a tool for the analysis of the distributed software evolution. According to this paper, collaborative frameworks and compatibility with collaboration will not only enable software developers to create applications in distributed teams but also help analysists work in (geographically) distributed frameworks to analyze the software evolution procedure. Such a tool will particularly be very efficient in open-source projects; hence, its user interface is Web-based and can be accessed at http://churrasco.inf.unisi.ch.

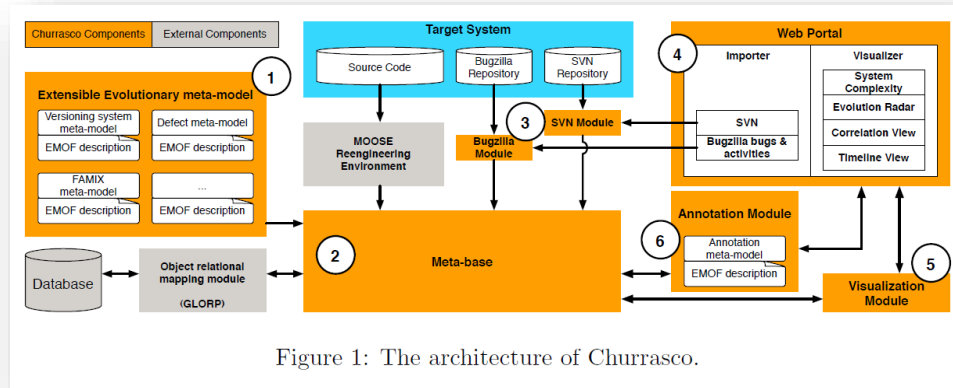

**Figure 4.** The architecture of Churrasco (Ref. [6])

This tool visualizes the analysis results in different formats (*e.g.*, a radar chart or a polymetric view).

An approach to the development process is to see it as a business process by considering a number of activities performed by a few individuals. It is possible to defined an order, a condition, a priority, a delay, and parallelism for these activities. Business processes can be modeled on activity networks (or other process modeling languages). A modeling process can also be executed.

The development process support systems use something similar to an activity network to model the development process and then execute the model.

However, what does a development process mean? It should be mentioned that the steps of a development process are usually taken by the project stakeholders and developers. When an activity is performed by someone, the process support system will be notified. Based on the process model and the current project status, the process support system determines the next process step and notifies those who should take the next step.

A coordinate of complex development processes (in addition to other coordinates such as the dynamism of changes in the development process, necessity of reusing products and objects, *etc.*) is the distribution of a process. In other words, the individuals and sub-processes gathered together in a general process should be able to interact within a distributed framework (in terms of time, location, *etc.*) to progress the software development process.

At the same time as the software evolution, the software development process should be completed in order to reflect the changes occurring along the software evolution in terms of dimension, complexity, domain, purpose, and other characteristics of a software application.

In a distributed project, *i.e.*, a project that is temporally or spatially distributed in its development process, the following tasks are very important and hard to accomplish:

- Establishing and maintaining constant communication between the project stakeholders (*e.g.*, developers)

- Coordinating the project stakeholders (*e.g.*, developers)
- Controlling project management

For this purpose, Ref. [14] proposed a system called *Endeavors* that provides the infrastructure for modeling and executing the distributed open development processes.

The following figure demonstrates an example of user interface in this system, the integration of process modeling and execution is important. The process model can be changed dynamically in this way to quickly reflect changes of the software evolution in the development process. In fact, this system integrates the development process modeling, administrative control of the process flow, and its execution.

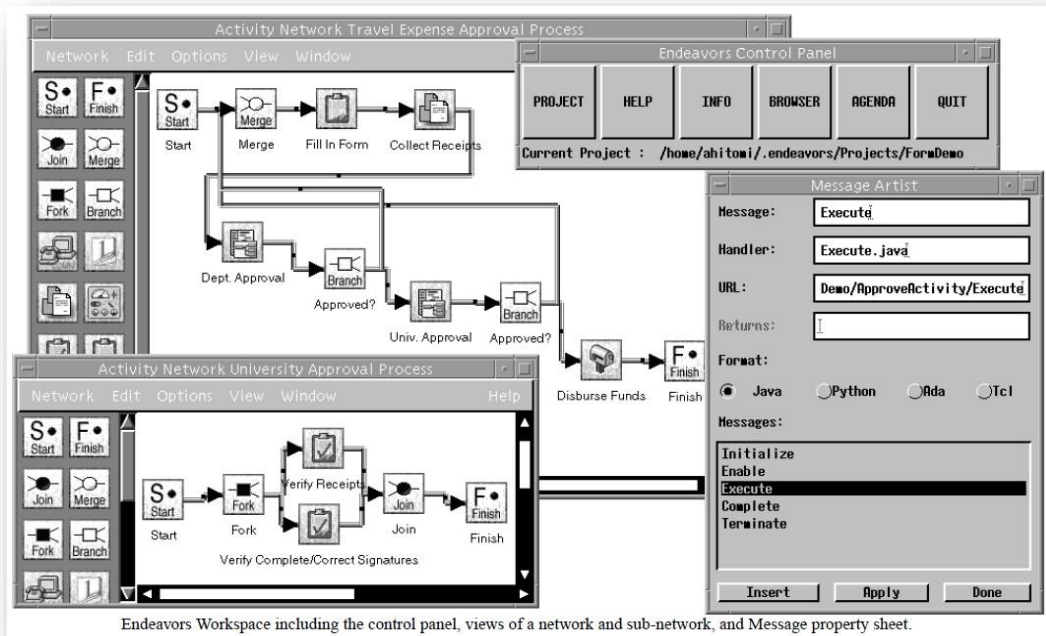

**Figure 5.** A schematic view of *Endeavors* [Ref. 14]

This system benefits from *Open Architecture* and a distribute development process. Its architecture has three levels: user, system, and foundation. The user level is responsible for facilitating the affairs of users and developers (by providing them with a "consistent view" of the system). Moreover, such patterns as the broad-casts and wrappers were employed to maintain the consistency of views. In the author's opinion, the system layer can be considered partly relevant to the implementation layer in *Open Architecture*, whereas the foundation layer can be considered relevant to its high-level architecture layer.

According to Ref. [7], if the software architecture does not provide the software constituents (*i.e.*, components and modules) with good independence, the dependence of constituents will prevent their distributed and paralleled development and evolution. Hence, the architectures of the software applications that are set to be developed globally or to evolve globally should support distribution in the development/evolution process. This doctoral dissertation pointed out how to achieve such an architecture whose support of development/evolution distribution would be guaranteed to some extent.

Moreover, collaboration was among the principles of this doctoral dissertation [2010], which defined collaboration as the support and establishment of communication between the members of the (geographically) distributed teams. Furthermore, distribution was introduced as a collaboration principle in this dissertation.

According to the literature review, collaboration is a keyword of distributed evolution because the distributed evolution and development teams now definitely need exchange and interaction.

Open-source software applications are developed within distributed frameworks. In other words, the individuals involved in the development of these applications are spatially and temporally distributed and act independently to some extent. Ref. [8] and its original reference addressed the effect of distribution of this evolution process on the coupling rate of the system constituents. Evidently, the coupling of the system components can completely affect the evolution of a system.

This empirical paper was published in 2003. According to the results of this paper on Apache Web Server, the distribution of a system evolution process cannot help conclude that the system has low or high evolutional capability. Some other papers tried to induce the theory that the distribution of a system evolution process would help that system have a low capability for higher evolution. The theory was rejected by this paper.

Ref. [22] addressed the challenge of the limited capability of interaction between the individuals involved in a project due to their physical distances. As a result, the project members will have insufficient knowledge about the project status. In this paper, the Palantir system was proposed to solve the abovementioned problem. This system visualizes the current changes in the project to inform the project members and demonstrate the effect and intensity of each change. In fact, it is a configuration management system that supports distribution, analyzes code changes, and uses visualization. According to the developers of the Palantir system, it can help establish close interactions between the project members. The authors of this paper believe that they integrated three areas of computer supported collaborative work (CSCW), configuration management, and software visualization.

By analyzing the changes, it is possible to determine how many changes occurred from the previous version of a document to its current version. The following techniques were employed to analyze the intensity of changes: the number of lines, token-based techniques, and abstract syntax techniques.

With the analysis of effect and effect domain, every individual can be informed about what potential effects each change has made on his/her working atmosphere and relevant code. The following techniques were employed to analyze the effect and effect domain of changes: the number of lines, the number of changed interfaces, and dependence analysis graph.

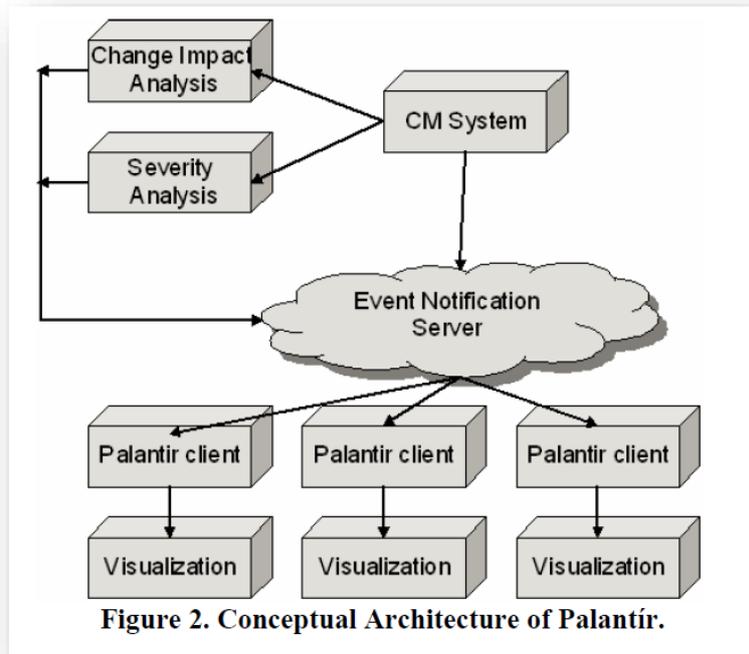

**Figure 6.** The conceptual architecture of Palantir (Ref. [22])

This empirical study (Ref. 23) indicated that dependences, especially logical dependences between software components and products, had important effects on software bugs. This paper also mentioned that logical dependences were not so clear, specific, and evident to developers. It also introduced some tools such as Palantir, TUKAN, and Ariadne that would provide the necessary knowledge for coordinating the project members. A suggestion of this paper, in addition to configuration management, is to inform the project members of changes, logical dependences, and other dependences of software components and products. Although Palantir performs this tasks to some extent through the analysis of change rate and change domain effect, this paper sought to provide more efficient knowledge. In fact, this paper tried to inform the project members of the project dependences, something which was identified as a requirement. In the distributed software development and evolution, emails and mail groups play a decisive, positive, and effective role in interactions and communications between the project members. Ref. [24] introduced Remail as a plugin for Eclipse to integrate the email transmission and reception features into the IDE environment and enable developers to interact and communicate with each other seamlessly within the development process via email. They can send the emails of source codes right near the source codes and interact with each other and exchange opinions on relevant subjects. According to this paper, the distributed development is not a common practice.

Ref. [26] conducted an empirical study on 189 distributed software projects. The distributed software projects that did not have a normal distribution (*i.e.*, there were more individuals centralized at some sites) were more prone to bugs that the projects with a normal distribution.

According to Ref. [27], there are certain challenges to cost estimation in the distributed software projects. Ref. [28] analyzed the effect of distribution of teams in the appropriateness of agile methodologies for projects, proposed some techniques for such teams, and mentioned different types of distributed teams. Furthermore, in the popular agile processes (*e.g.*, XP and Scrum), the methodology monitors the entire software process in both the initial development step and the software evolution step. Although all aspects of the software evolution are not covered in these methodologies, the important point is that most of the agile processes, but not all of them, emphasize that the activities of all team members should be at the same time and location. Hence, many of such methodologies cannot suit the distributed software projects, and only some of them are usable for distributed projects in terms of distribution and process calibration. However, the "agile global software development" is a well-known term in the literature (*e.g.*, refer to Ref. [29]). This term comes from the combination of two concepts, *i.e.*, software project distribution and process agility.

Many global software projects are characterized by the concepts of "original site" and "distributed sites"; however, these projects are faced with one problem. It is the distrust of the original site in the quality and tasks of the distributed sites as well as the negative attitudes of the original site members towards the members of distributed sites (Ref. [30]).

From a theoretical standpoint, dependence on the project centrality reduces flexibility and causes some problems in the distributed projects that are centralized (*i.e.*, they are distributed but are not decentralized. For instance, they only use the distributed development teams, outsourcing, or off-site development while the main development and evolution process is directed and navigated within a centralized framework). However, this centrality has some positive effects on the centralized and distributed projects. For instance, it appears that it is easier to direct and control the project process accurately in centralized projects than in decentralized projects. Furthermore, it can be easier to maintain the integrity and consistency of products and applications in centralized projects than some decentralized projects. Practically, it is easier and more feasible to monitor and audit the quality of centralized projects than that of decentralized ones. However, some attempts have been made to eliminate these differences between centralized and decentralized projects (*e.g.*, proposing *Open Architecture* by Arizi).

Many of the successful FLOSS projects (representing a specific type of decentralized distributed projects) are practically characterized by a centralized interactive agent (*e.g.*, the project founder or an individual who affects the project much more than the project members by literally being concerned to direct the project or some individuals who have created a consistent coordinated central group for the project via email and mail groups). Hence, the claim that the successful FLOSS projects are characterized by a high level of decentralization should be hesitated and should not be accepted easily. However, there are always some exceptions.

In the distributed software evolution, the project members form a social network in practice, and their other physical dependences (*e.g.*, time, location, *etc.*) fade away. Hence, a research opportunity in this area is to regard the project members as a social network and analyze their collective behavior as the "outcome" of a social network (Ref. [34]). The common denominator of distributed evolution and social networks leads to a rich bulk of papers.

There are some tools for establishing communications and interactions between distributed teams. They are used for specific documents and products. For instance, Requirement Explorer was designed and proposed to report, share, and meet requirements by distributed teams (Ref. [35]).

According to Ref. [39], the project member should be informed of the "instabilities" of a project in a collectively distributed project, like what occurs in a distributed software project. In fact, everyone should have common knowledge about these instabilities. In other words, the "inconsistencies" of products and "distrusts and doubts" in a project should be considered in the project decision-making process. It can be concluded from this paper that the collective knowledge in a distributed project should not be limited to certain aspects of the project. It would not be enough to inform everyone of what sections of the project have been finalized. However, everyone should be made aware of defects, inconsistencies, distrusts, and even doubts. In other words, the knowledge about a project is not a type of limited knowledge. It is more inclusive, something which can be related to the characteristic of "distributed cognition" because the prerequisite to distributed cognition is the availability of knowledge to the project members with regard to certain aspects such as inconsistencies and uncertainties. Nevertheless, despite the distribution of a project, it will be centralized in terms of cognition and knowledge at the very least, something will result in the problems previously mentioned for centralized projects. However, decentralization is not suitable for every project. For some, centralization is necessary. In particular, if the "distributed cognition" is not appropriate or not possible, the decentralization of project will not be a good solution. In this case, the centralized distribution of a project can be beneficial.